\documentclass[12pt]{article}
\usepackage[sumlimits,intlimits,namelimits]{amsmath}
\usepackage{amssymb}
\usepackage[english]{babel}
\usepackage{bbm}
\usepackage{parskip}
\global\parskip 6pt

\newcommand{\be}{\begin{eqnarray}}
\newcommand{\ee}{\end{eqnarray}}
\newcommand{\non}{\nonumber}
\newcommand{\lb}{\label}
\def\>{\rangle}
\def\<{\langle}

\begin{document}
%
\begin{titlepage}
\hfill{LMU-ASC 47/10 }
\vspace*{.5cm}
\begin{center}
{\Large{{\bf Classical Stability of the BTZ Black Hole \\[.5ex]
in Topologically Massive Gravity}}} \\[.5ex]
\vspace{1cm}
Danny Birmingham\footnote{Email: dbirmingham@pacific.edu}\\
\vspace{.1cm}
{\em Department of Physics,\\
University of the Pacific,\\
Stockton, CA 95211\\
USA}\\
\vspace*{.5cm}
Susan Mokhtari\footnote{Email:
susan@science.csustan.edu}\\
\vspace{.1cm} {\em Department of Physics,\\
California State University Stanislaus,\\
Turlock, CA 95380,\\
USA} \\
\vspace*{.5cm}
Ivo Sachs\footnote{Email: ivo.sachs@physik.uni-muenchen.de}\\
{\em Arnold Sommerfeld Center,\\
Ludwig-Maximilians Universit\"{a}t,\\
Theresienstrasse 37, D-80333, M\"{u}nchen,\\
Germany}\\
\end{center}
\vspace*{1cm}
\begin{abstract}
\noindent {We demonstrate the classical stability of the BTZ black
hole within the context of topologically massive gravity. The
linearized perturbation equations can be solved exactly in this
case. By choosing  standard  boundary conditions appropriate to the stability
problem, we demonstrate the absence of modes which grow
in time, for all values of the Chern-Simons coupling.}
\\
\vspace*{.25cm}
\end{abstract}
\vspace*{.25cm}
\end{titlepage}
\section{Introduction}

Topologically massive gravity (TMG) \cite{Deser:1982vy},  in three-dimensional anti-de Sitter space has been the subject of considerable
interest recently (see, for example, \cite{Witten07}-\cite{Grumiller:2009mw} for a partial list of references). As in any theory, a first step is to search for solutions to the classical equations of motion.
Fortunately, the structure of the equations for TMG guarantee that the known solutions
without a Chern-Simons coupling  $\mu =\infty$  are also solutions when $\mu $ is finite.
In particular, then, we already have a constant curvature black hole solution at our disposal, namely
the BTZ black hole \cite{BTZ}. One of the most remarkable features of the BTZ black hole is the role that it has played
in understanding many aspects of the AdS/CFT correspondence, and its role in the near-horizon geometry of higher-dimensional black holes.

Given a black hole solution, a first order of business is to examine its classical stability properties. Typically, this is
accomplished by resorting to the linearized approximation, and exploring solutions to the associated
boundary value problem. Our purpose here is to examine the stability of the BTZ black hole within the context of
topologically massive gravity, for all values of the Chern-Simons coupling $\mu$.

The key to our analysis lies in the fact that the perturbation
equations can be solved exactly. We can thus search explicitly for
modes which grow in time; the presence of such modes
would indicate a classical instability. Of course, a crucial
ingredient in this analysis is the choice of boundary conditions.
The original asymptotic boundary conditions for three-dimensional
anti-de Sitter gravity were determined by Brown and Henneaux \cite{BH}.
Recently, it has been shown that one may relax these conditions
slightly in topologically massive gravity, for certain ranges of the
coupling $\mu$ \cite{Henneaux:2009pw}. We adopt these generalized 
boundary
conditions at asymptotic infinity as the appropriate conditions to impose on the linear
perturbations. In the presence of a black hole background, one also
needs to impose boundary conditions at the horizon. We  establish
the fact that by simply demanding boundedness of the perturbation
(necessary for the linearized approximation to be valid), the
absence of unstable modes is guaranteed. Stability of black holes in warped $AdS$ has recently been discussed in \cite{Anninos:2010pm}.

The plan of this paper is as follows. In section 2, we present the basic equations of topologically massive gravity and solve directly the first order equations of motion in the BTZ background. 
In section 3, we identify
potential unstable modes and confront them with the appropriate boundary conditions. In section 4, we present an alternative approach to the problem, which is based on the second order analysis
of \cite{Carlip:2008jk}. The absence of unstable modes in also confirmed from this viewpoint.
In section 5, we conclude with some brief remarks.

\section{ Metric Perturbations}

The action for topologically massive gravity is taken in the form
 \be
S&=&-\frac{1}{16\pi G}\int d^{3}x\sqrt{-g}\left(R+\frac{2}{l^2}\right)-\frac{1}{32\mu
\pi G}\int d^{3}x\sqrt{-g}\epsilon^{\lambda\mu\nu}\Gamma^{\rho}_{\lambda\sigma}
\left(\partial_{\mu}\Gamma^{\sigma}_{\nu\rho} +\frac{2}{3}\Gamma^{\sigma}_{\mu\tau}
\Gamma^{\tau}_{\nu\rho}\right),\lb{action}
\ee
where $\mu$ is the Chern-Simons coupling, and the parameter $l$ sets the scale of the cosmological constant of anti-de Sitter space,
$\Lambda = -1/l^{2}$. 

In the following, we will be interested in the linear approximation around a background spacetime, and therefore decompose
the metric tensor as $g_{\mu\nu} \rightarrow g_{\mu\nu} + h_{\mu\nu}$. Working in the transverse traceless gauge, with
$\nabla^{\mu}h_{\mu\nu} = g^{\mu\nu}h_{\mu\nu} = 0$, the equation of motion takes the form
\be
(\nabla^{2} + 2)[h_{\mu \nu} + \frac{1}{\mu} \epsilon_{\mu}^{\phantom{\mu}\alpha\beta}\nabla_{\alpha}h_{\beta\nu}] = 0.
\ee
It is convenient to define the operators
\be
({\cal D}^{M})_{\mu}^{\phantom{\mu}\beta} = \delta_{\mu}^{\phantom{\mu}\beta} + \frac{1}{\mu}\epsilon_{\mu}^{\phantom{\mu}\alpha\beta}\nabla_{\alpha},
\;\;
({\cal D}^{L/R})_{\mu}^{\phantom{\mu}\beta} = \delta_{\mu}^{\phantom{\mu}\beta} \pm l\epsilon_{\mu}^{\phantom{\mu}\alpha\beta}\nabla_{\alpha}.
\ee
The third order equation of motion can then be written in the form \cite{Strominger}
\be
({\cal D}^{L}{\cal D}^{R}{\cal D}^{M})_{\mu\nu} = 0.
\ee%
The first order equation for a massive graviton is given by $({\cal D}^{M}h)_{\mu\nu} = 0$, namely
\begin{equation}\label{first}
(\epsilon_\mu^{\;\;\alpha\beta}\nabla_\alpha h_{\beta\nu}+\mu)
h_{\mu\nu}=0~~.
\end{equation}
The BTZ black hole metric can be written in the form
\be\label{guv}
ds^{2} = -\sinh^{2}\rho\; dt^{2} + \cosh^{2}\rho \;
d\phi^{2} + d\rho^{2},
\ee
where we have introduced the radial coordinate $r = \cosh \rho$; the horizon and infinity then correspond to $\rho = 0, \infty$, respectively.
We choose units such that the mass of the black hole is $M=1$ and set $l=1$. In the following,
we will also use coordinates $u = t + \phi$ and $v = t - \phi$.

To solve the equation of motion (\ref{first}), we make an ansatz for the perturbation in the form
\be
h_{\mu\nu} = e^{-i\omega t -ik\phi}\left(\begin{array}{ccc}
F_{uu} &F_{uv}&F_{u\rho}\\
F_{vu}&F_{vv}&F_{v\rho}\\
F_{\rho u}& F_{\rho v}& F_{\rho\rho}\end{array}\right).
\ee
Using $\epsilon^{\rho u v}={1\over \sqrt{-g}}={4\over \sinh 2\rho}$, the equations of motion can be written in the form \cite{Grumiller:2009mw}
\be
\bar{h} F_{uu} - h F_{uv} &=& \left(\frac{-\mu -1}{4i}\right) \sinh (2\rho) F_{u\rho},\label{eqn1}\\
\bar{h} F_{uv} - h F_{vv} &=& \left(\frac{-\mu +1}{4i}\right) \sinh (2\rho) F_{v\rho},\label{eqn2}\\
\bar{h} F_{u\rho} - h F_{v\rho} &=& \frac{i}{\sinh(2\rho)}[-F_{vv}(-\mu+1) - F_{uu}(-\mu-1) + 2\mu\;\cosh(2\rho)F_{uv}],\label{eqn3}\\
F_{\rho\rho} &=&  \frac{4}{\sinh^{2}(2\rho)}[2\cosh(2\rho)F_{uv} + F_{uu} + F_{vv}],\label{eqn4}\\
\non
\ee
together with the differential equations
\be
\frac{dF_{uv}}{d\rho} &=& \left(\frac{-\mu + 1}{\sinh(2\rho)}\right)\left[F_{uv}\left(\frac{4h\bar{h}}{(-\mu+1)^{2}} - \cosh(2\rho)\right)
-F_{vv}\left(1 + \frac{4h^{2}}{(-\mu+1)^{2}}\right)\right],\label{eqn5}\\
\frac{dF_{vv}}{d\rho} &=& \left(\frac{-\mu + 1}{\sinh(2\rho)}\right)\left[- F_{vv}\left(\frac{4h\bar{h}}{(-\mu+1)^{2}} - \cosh(2\rho)\right)
+F_{uv}\left(1 + \frac{4\bar{h}^{2}}{(-\mu+1)^{2}}\right)\right]\label{eqn6},
\ee
where we have defined $h = (\omega + k)/2, \bar{h} = (\omega - k)/2$. The case of chiral gravity $\mu \pm 1$ will be treated separately
below.

The equations (\ref{eqn5}) and (\ref{eqn6}) are readily transformed into a second order hypergeometric equation
(see \cite{Grumiller:2009mw} for details).
The asymptotic $\rho$-dependence of the two solutions is then given by
$F_{vv}\propto e^\rho$ and $F_{vv}\propto e^{(1-\mu)\rho}$, respectively.
Since the former solution is excluded by the asymptotic boundary conditions given below, we shall focus attention on the latter.
A particular solution of (\ref{eqn1})-(\ref{eqn6})
is found by setting $F_{uu} = F_{uv} = F_{u\rho} =0$. We then have  
\be
F_{v\rho} &=& \frac{i}{\sinh(2\rho)}\left(\frac{-\mu +1}{h}\right)F_{vv},\\
F_{\rho\rho} &=& \frac{4}{\sinh^{2}(2\rho)}F_{vv},\\
h &=& \pm \frac{i}{2}(-\mu + 1).
\ee
Choosing the branch $h = \frac{i}{2}(-\mu + 1)$, we find a right-moving solution
\be\label{hr}
h^{R}_{\mu\nu}= e^{(1 - \mu)t +ik(t-\phi)}(\sinh \rho)^{1 -\mu} (\tanh \rho)^{ik}
\left(\begin{array}{ccc}
0&0&0\\
0&1&\frac{2}{\sinh(2\rho)}\\
0&\frac{2}{\sinh(2\rho)}&\frac{4}{\sinh^{2}(2\rho)}\end{array}\right),
\ee
while  the branch $h = -\frac{i}{2}(-\mu + 1)$ also leads to a right-moving solution
\be\label{Hr}
H^{R}_{\mu\nu}= e^{(\mu -1)t +ik(t-\phi)}(\sinh \rho)^{1 - \mu} (\tanh \rho)^{-ik}
\left(\begin{array}{ccc}
0&0&0\\
0&1&-\frac{2}{\sinh(2\rho)}\\
0&-\frac{2}{\sinh(2\rho)}&\frac{4}{\sinh^{2}(2\rho)}\end{array}\right).
\ee
From the above expressions we see explicitly that   (\ref{hr}) and (\ref{Hr})  are ingoing and outgoing at the horizon, respectively.
The outgoing modes are relevant for white holes while the ingoing modes are relevant for black
holes\footnote{In fact, it turns out that even if we where to keep the outgoing modes
they would be eliminated by the boundary conditions given below.}. We will thus focus on the mode  (\ref{hr}).

There is another class of ingoing
solutions  generated by making
the replacements $u\leftrightarrow v, h\leftrightarrow \bar{h}, \mu \rightarrow -\mu$, leading to
\be\label{hl}
h^{L}_{\mu\nu}= e^{(1 +\mu)t -ik(t+\phi)}(\sinh \rho)^{1 +\mu} (\tanh \rho)^{-ik}
\left(\begin{array}{ccc}
1&0&\frac{2}{\sinh(2\rho)}\\
0&0&0\\
\frac{2}{\sinh(2\rho)}&0& \frac{4}{\sinh^{2}(2\rho)}\end{array}\right).
\ee

\section{ Stability}
The task now is to identify those solutions which obey the generalized boundary conditions \cite{Henneaux:2009pw} at infinity, and which grow
in time. In terms of the coordinates $(\rho,u,v)$, an admissible metric perturbation has to satisfy
either the
boundary conditions
\be\label{bn}
h_{\rho\rho} &=& e^{-2\rho}f_{\rho\rho},\non\\
h_{\rho u} &=& e^{-2\rho}f_{\rho u},\non\\
h_{\rho v} &=& k_{\rho v}\;e^{-(1+\mu)\rho} + e^{-2\rho} f_{\rho v},\non\\
h_{uu} &=& f_{uu},\non\\
h_{uv} &=& f_{uv},\non\\
h_{vv} &=& k_{vv} \;e^{(1-\mu)\rho} + f_{vv},
\ee or the
boundary conditions
\be\label{br}
h_{\rho\rho} &=& e^{-2\rho}f_{\rho\rho},\non\\
h_{\rho u} &=& k_{\rho u}\;e^{(-1 + \mu)\rho} + e^{-2\rho}f_{\rho u},\non\\
h_{\rho v} &=& e^{-2\rho}f_{\rho v},\non\\
h_{uu} &=& k_{uu} \;e^{(1+\mu)\rho} + f_{uu},\non\\
h_{uv} &=& f_{uv},\non\\
h_{vv} &=& f_{vv}.
\ee
The additional $\mu$-dependent terms are only present in the range $\mid\!\mu\!\mid\; <1$.
Upon examination of the above solutions, we
observe that the ingoing solutions grow exponentially in time, and
also satisfy these boundary conditions, provided $\mid\!\mu\!\mid\;<1$.
These solutions thus represent potentially unstable modes. The right-moving perturbation
$h^{R}_{\mu\nu}$ obeys the boundary conditions (\ref{bn}), while the left-moving solution
$h^{L}_{\mu\nu}$  obeys the boundary conditions (\ref{br}). We wish to determine if such solutions can be eliminated
by imposing a physically acceptable boundary condition at the horizon.
To see this, it suffices to invoke boundedness of the solution, which is required in order for
the linear approximation to be valid \cite{Gibbons}-\cite{Gregory}. Boundedness of the solution near the horizon
is most easily seen by transforming to Kruskal
coordinates \cite{Teitelboim}
\be
R = \tanh \frac{\rho}{2}\;\cosh t,\;\; T = \tanh \frac{\rho}{2}\;\sinh t.
\label{kruskal}
\ee
Since the Kruskal coordinates are well defined at the horizon, we must also require the perturbation to be well behaved there.
However, one can check that the Kruskal components of $h^{L}_{\mu\nu}$  and $h^{R}_{\mu\nu}$ diverge at the horizon,
thus excluding them as physically acceptable.
Alternatively, one may state that boundedness of the perturbation requires the existence of a non-singular
linearized diffeomorphism,   such that $g^{\mu\alpha}h_{\alpha\nu} <<1$, and this leads to boundary conditions
both at the horizon and infinity.
%
Given the explicit expression
\begin{equation}
(h^L)^\mu_{\;\nu}= e^{(1 +\mu)t -ik(t+\phi)}(\sinh \rho)^{1 +\mu} (\tanh \rho)^{-ik}\begin{pmatrix}
        -\frac{4}{\sinh^2(2\rho)}&0&-\frac{8}{\sinh^3(2\rho)}\cr
        -\frac{4\cosh(2\rho)}{\sinh^2(2\rho)}&0&-\frac{8\cosh(2\rho)}{\sinh^3(2\rho)}\cr
        \frac{2}{\sinh(2\rho)}&0&\frac{4}{\sinh^2(2\rho)}
        \end{pmatrix},
\end{equation}
which satisfies the asymptotic boundary conditions at infinity for $|\mu|<1$, it is not difficult to show that a
non-singular  diffeomorphism that renders the perturbation bounded at the horizon cannot exist for
this range of $\mu$. A similar argument applies to the right-moving perturbation $(h^R)^\mu_{\;\nu}$.
Thus,  the solutions are excluded as physically unacceptable.

The generic metric perturbation with the same asymptotic behaviour as (\ref{hl})
has a time dependence given by replacing $\mu\to \mu-2n$, $n\in{\bf N}$ in the
exponent \cite{SS08}. However, for $n\neq 0$, there are no growing modes which satisfy the asymptotic boundary
conditions.  Similarly, generic metric perturbation with the same asymptotic behaviour as (\ref{hr}) have a time
dependence given by replacing $\mu\to \mu+2n$, $n\in{\bf N}$.
Again, such modes are excluded by the  asymptotic boundary conditions.

The remaining case to deal with is when $\mu = \pm 1$. The modes
(\ref{hr}) and (\ref{hl}) then become pure gauge transformations and thus do not represent a physical perturbation of the black hole.
However, at the chiral point $\mu=1$, a new class of logarithmic modes arises \cite{Grumiller:2008qz}.
It is obtained by differentiating (\ref{hr}) before setting $\mu=1$ \cite{ivo}, leading to a solution of the form
\be
 h^{log}_{\mu\nu}=-y(\tau,\rho)h^R_{\mu\nu}\;,\lb{n0}
\ee
where $y(\tau,\rho)=\tau+\ln[\sinh(\rho)]$. The generalized boundary conditions for $\mu = \pm 1$ can be obtained
by making the replacement $e^{(1-\mu)\rho} \rightarrow \rho$ in (\ref{bn}),  and $e^{(1+\mu)\rho} \rightarrow \rho$ in (\ref{br}).
The asymptotic $\rho$-dependence of
(\ref{n0}) is then consistent with these generalized boundary conditions. However, as is clear from the form
of $y(\tau,\rho)$, the non-boundedness of $h^R_{\mu\nu}$ at the horizon implies the
non-boundedness of $h^{log}_{\mu\nu}$, and consequently all of its descendants. The anti-chiral point $\mu=-1$ is treated in a similar fashion.

In conclusion, we have show that all potentially unstable solutions, growing in time and obeying the generalized
boundary conditions at asymptotic infinity are excluded by the requirement of boundedness of the solution at the horizon.
According to these criteria, the BTZ black hole is thus a stable solution of topologically massive gravity for all
values of the Chern-Simons coupling $\mu$.

\section {Scalar Formulation of Metric Perturbations}
In \cite{Carlip:2008jk}, it was shown that the action for topologically
massive gravity can be re-cast in terms of a single massive scalar
field, with the mass related to the Chern-Simons coupling parameter.
Consequently, it was established that the perturbation
equations for all gauge invariant modes can be formulated as second order massive
scalar field equations.
It is well know that the equation for a massive scalar field in the background of the BTZ metric can be solved exactly in terms of hypergeometric
functions, and this will allow us to explicitly study the stability of the BTZ black hole.
As well as confirming the analysis of the previous section, it will also highlight an alternative viewpoint on the boundary conditions
relevant for a stability analysis.

It is well known that there exists a class of topological black holes in anti-de Sitter
space, with line element
\cite{Birtopbh}
\be
ds^{2} = - f(r)\; dt^{2} + f^{-1}(r)\;dr^{2} +
r^{2}h_{ij}(x)\;dx^{i}dx^{j}, \label{topbh}
\ee
where
\be f(r) =
\left(k - \frac{2M}{r^{d-3}} +\frac{r^{2}}{l^{2}}\right).
\ee
The parameter $k$ can take the
values $k=1, 0, -1$, and
the cosmological
constant is $\Lambda = -(d-1)(d-2)/2l^{2}$.
The novel feature of these topological black holes is the fact that
there exists a massless black hole when $k=-1$.
The crucial point to note here is that the metric ansatz for the BTZ black hole (with mass parameter equal to one)
is of the same form as the massless topological black hole with $M=0, k=-1, d=3$, and we set $l=1$.
Thus,
the stability analysis of the massless topological black hole performed in \cite{BM} can be used to analyze
the stability of the BTZ black hole within the context of topologically massive gravity.
In \cite{BM}, the massive scalar field equation was solved exactly.
However, in order to apply those results to the case at hand, we
need to specify the scalar field in terms of metric components,
and re-analyze the stability question within that context.

To proceed, we consider a scalar field $\phi$ of mass $m$ in the BTZ background,
\be
(\nabla^{2} - m^{2})\phi =0.
\ee
As shown in \cite{Carlip:2008jk}, the essential dynamics of topologically massive gravity is encoded in
a scalar field of mass $m^{2} = (\mu + 2)^{2} -1$. Furthermore, all independent gauge invariant perturbations
can be re-cast as scalar field equations for various masses. Choosing the ansatz
\be
\phi =
\phi(r)e^{\omega t -ik\phi},
\ee
brings the radial equation to the
form
\be
\left[-\left(f\frac{d}{dr}\right)^{2} + V\right]\Phi(r) = -
\omega^{2}\Phi(r), \label{master}
\ee
where $\Phi = r^{\frac{1}{2}}\phi$, and
\be
V = \frac{f}{r^{2}}\left[k^{2} + \frac{1}{4} +
\left(\frac{3}{4} +
m^{2}\right)r^{2}\right]. \label{Vfield}
\ee

In order to investigate the stability properties of the black hole,
it is useful to recast Eq. (\ref{master}) as a Sturm-Liouville
equation
\be
A\Phi =
\lambda \Phi, \label{Aoperator1}
\ee
where the Schr\"{o}dinger operator is given by
\be
A =
-\frac{d^{2}}{dr_{*}^{2}} + V(r),  \label{Aoperator2}
\ee
with eigenvalue $\lambda = -\omega^{2}$, and the
tortoise coordinate $r_{*}$ is defined by $dr_{*} = \frac{dr}{f}$.
Our task is to solve this equation subject to appropriate boundary
conditions. Given the Sturm-Liouville form, this involves searching
for eigenvalue solutions which are normalizable with respect to the standard measure
\cite{Gibbons, Kodama2},
\be 1 = \int dr_{*}\;\Phi^{*}\Phi.
\label{norm}
\ee
In particular, unstable modes will correspond to normalizable
($\omega > 0$) states of the Schr\"{o}dinger operator $A$.

Near the horizon, this condition of normalizability demands that the
solution behave as $\Phi \sim (r-1)^{\epsilon}$, and thus we impose
a Dirichlet boundary condition $\Phi \rightarrow 0$ on the
perturbation \cite {Gibbons, Kodama2, KayWald}. For large $r$,
normalizability requires $\Phi \sim r^{\frac{1}{2} - \epsilon}$.
However, the scalar field for TMG is related to a metric component
by \cite{Carlip:2008jk} \be \Phi = z^{3/2}h_{zz}, \ee where
the upper half-space coordinate $z \sim
\frac{1}{r}$ for large $r$. In terms of
this coordinate, the generalized asymptotic boundary condition is
$h_{zz} = O(1)$, and thus we require $\Phi \sim \frac{1}{r^{3/2}}$,
for large $r$.

To proceed towards the solution of (\ref{master}), we change
variables to a new radial coordinate defined by
\be z = 1 - \frac{1}{r^{2}}. \ee Thus, $z=0$ now
corresponds to the location of the horizon $r = 1$, while $z=1$
corresponds to $r=\infty$. The master equation can then be written
as
\be z(1-z)\frac{d^{2}\Phi}{dz^{2}} + \left(1 -
\frac{3z}{2}\right)\frac{d\Phi}{dz} +\left[\frac{A}{z} + B +
\frac{C}{1-z}\right]\Phi = 0, \ee
where
\be
A = -\frac{\omega^{2}}{4},\;\;
B = -\frac{1}{4}\left(\frac{1}{4} + k^{2}\right),\;\;
C = -\frac{1}{4}\left(m^{2} + \frac{3}{4}\right).
\ee
Defining
\be
\Phi(z) = z^{\alpha}(1-z)^{\beta}F(z),
\ee
allows the
master equation to be reduced to hypergeometric form
\be
z(1-z)\frac{d^{2}F}{dz^{2}} + [c -
(a+b+1)z]\frac{dF}{dz} - ab F = 0, \label{hyper}
\ee provided that
\be
\alpha = \pm\frac{\omega}{2},\;\;
\beta = \frac{1}{4} \pm \frac{1}{2}\sqrt{1 + m^{2}},
\label{beta}
\ee
with the coefficients determined as followed
\be
a &=& \frac{1}{4} + \alpha + \beta  + \frac{1}{2}\sqrt{- k^{2}}, \non\\
b &=& \frac{1}{4} + \alpha + \beta - \frac{1}{2}\sqrt{- k^{2}},\non\\
c &=& 2 \alpha + 1. \label{abc}
\ee
Without loss of generality, we
can take
\be
\alpha =  \frac{\omega}{2},\;\;
\beta = \frac{1}{4} - \frac{1}{2}\sqrt{1 + m^{2}}.
\label{alphabeta}
\ee

In the neighbourhood of the horizon, the two linearly independent
solutions of (\ref{hyper}) are $F(a,b,c,z)$ and $z^{1-c}F(a-c+1,
b-c+1,2-c,z)$. With the choice (\ref{alphabeta}), the solution which
is regular (satisfying Dirichlet boundary conditions) at the horizon
is then given by
\be
\Phi(z) = z^{\alpha}(1-z)^{\beta}F(a,b,c,z).
\ee
Having imposed the Dirichlet boundary condition at the horizon,
we can now analytically continue this solution to infinity. In
general, the form of the solution near $z=1$ is given by
\cite{Abram}
\be
\Phi &=& z^{\alpha}(1-z)^{\beta}\frac{\Gamma(c)\Gamma(c-a-b)}{\Gamma(c-a)\Gamma(c-b)}F(a,b,a+b-c+1,1-z)\non\\
&+& z^{\alpha}(1-z)^{\beta +
c-a-b}\frac{\Gamma(c)\Gamma(a+b-c)}{\Gamma(a)\Gamma(b)}F(c-a, c-b,
c-a-b+1, 1-z), \label{infty}
\ee
where $c-a-b = \frac{1}{2} - 2 \beta$.
The generalized asymptotic boundary condition requires that $\Phi \sim (1 - z)^{3/4}$ near $z=1$.
First, we consider the case when $m^{2} >0$. Then $\beta <-\frac{1}{4}$ and the second term in
(\ref{infty}) clearly vanishes at infinity. In order to guarantee the vanishing of the
divergent first term, we must demand that
\be
c - a = -n, \;\; or \;\;c-b = -n,
\label{ca}
\ee
where $(n=0,1,2,3,...)$. In particular, the condition $c-a = -n$ becomes
\be
\omega = -1 - \sqrt{1 + m^{2}} + \sqrt{-k^{2}} -2n.
\ee
It is then clear that unstable modes with $\omega >0$ do not exist since $k^{2} \geq 0$.

For $m^{2} = 0$, we have $\beta = -1/4$, and $c-a-b = 1$. As a result, the analytic continuation to $z=1$
contains logarithmically divergent terms, and is given by
\be
\Phi &=& z^{\alpha}(1-z)^{-1/4}\frac{\Gamma(a+b+1)}{\Gamma(a+1)\Gamma(b+1)}\non\\
&+&
\frac{\Gamma(a+b+1)}{\Gamma(a)\Gamma(b)}z^{\alpha}(1-z)^{3/4}\sum_{n=0}^{\infty}\frac{(a+1)_{n}(b+1)_{n}}{n!(n+1)!}
(1-z)^{n}[{\mathrm{ln}}(1-z) \non\\
&-& \psi(n+1) -\psi(n+2) + \psi(a+n+1) + \psi(b+n+1)],
\label{5d}
\ee
where $(a)_{n} = \Gamma(a+n)/\Gamma(a)$, and $\psi(z) = \Gamma^{\prime}(z)/\Gamma(z)$.
To guarantee the absence of the divergent first term, we now require $a+1 = -n$ or $b+1 = -n$. Note that these conditions
also ensure the vanishing of the logarithmic terms in (\ref{5d}). Since $c-a-b=1$, we can write
these conditions as (\ref{ca}), which we have already shown have no solutions.

For $-1 <m^{2} < 0$, the range of $\beta$ is $-\frac{1}{4} < \beta < \frac{1}{4}$, and the solution
is given by (\ref{infty}). The absence of unstable solutions is again guaranteed by (\ref{ca}).

Finally, for $m^{2} = -1$, we have $\beta = \frac{1}{4}$, and $c-a-b = 0$. The solution then
takes the form
\be
\Phi &=& z^{\alpha}(1-z)^{1/4} \frac{\Gamma(a+b)}{\Gamma(a)\Gamma(b)}\sum_{n=0}^{\infty}
\frac{(a)_{n}(b)_{n}}{(n!)^{2}}[2\psi(n+1)
- \psi(a+n) \non\\
&-&\psi(b+n) - {\mathrm{ln}}(1-z)](1-z)^{n}.\non\\
\ee
Consistency with the generalized asymptotic boundary conditions requires $a=-n$ or $b =-n$. However,
since $c-a-b = 0$, these conditions again reduce to (\ref{ca}).

In conclusion, we have established the absence of unstable modes for the BTZ black hole within the scalar field
formulation of topologically gravity. This result confirms the first order analysis of the previous section.
In the previous section, we used boundedness of the perturbation at the horizon to eliminate the potentially
unstable modes. It is worth mentioning that these modes can also be eliminated by
requiring the perturbation to be normalizable at the horizon.
Normalizability at the horizon requires that $\Phi \sim (r-1)^{\epsilon}$,
and hence
$h_{\rho\rho}\sim \rho^{2 + \epsilon}.$
By examining the solutions  $h^{L}$ we see that normalizability at the horizon requires $\mu > 3$.
For $h^{R}$ normalizability at the horizon requires $\mu <-3$. Both conditions are incompatible with
the generalized asymptotic boundary conditions, thereby excluding such modes.

\section {Discussion}
We have discussed the classical stability of the BTZ black hole as a solution of topologically massive gravity.
The linearized perturbation equations are exactly solvable, and this allowed us to explicitly
examine the behaviour of the solutions subject to certain boundary conditions at the horizon and infinity.
Using the Brown-Henneaux boundary conditions at infinity, extended to incorporate the Chern-Simons term,
and boundedness of the perturbation at the horizon, allows us to establish the stability of the BTZ
black hole. This result was confirmed by studying the perturbation within the scalar field
formulation of \cite{Carlip:2008jk}, and using normalizability of the perturbation.\\ \\

\noindent {\Large \bf  Acknowledgments}\\ \\
D.B and S.M are supported by the National Science Foundation under grant PHY-0855133.
I.S was supported in part by the Transregional Collaborative Research Centre TRR 33,
the DFG cluster of excellence ``Origin and Structure of the Universe'' as well as the DFG project Ma 2322/3-1.
I.S would like to thank the University of the Pacific for hospitality and financial support during part of the work and V. Mukhanov for discussions.


\begin{thebibliography}{99}

\bibitem{Deser:1982vy}
  S.~Deser, R.~Jackiw and S.~Templeton,
  Phys.\ Rev.\ Lett.\  {\bf 48} (1982) 975; S.~Deser, R.~Jackiw and S.~Templeton,
  Annals Phys.\  {\bf 140} (1982) 372.

\bibitem{Witten07}
  E.~Witten,
  [arXiv:0706.3359 [hep-th][.


\bibitem{Strominger}
  W.~Li, W.~Song and A.~Strominger,
  JHEP {\bf 0804}, 082 (2008)
  [arXiv:0801.4566 [hep-th]].

\bibitem{Carlip:2008jk}
  S.~Carlip, S.~Deser, A.~Waldron and D.~K.~Wise, Class. Quant. Grav. 26 (2009) 075008
  [arXiv:0803.3998 [hep-th]].

\bibitem{Grumiller:2008qz}
  D.~Grumiller and N.~Johansson,
  [arXiv:0805.2610 [hep-th][.

\bibitem{SS08}
  I.~Sachs and S.~N.~Solodukhin,
  JHEP {\bf 0808} (2008) 003
  [arXiv:0806.1788 [hep-th]].

\bibitem{Grumiller:2009mw}
  D.~Grumiller and I.~Sachs,
  JHEP {\bf 1003} (2010) 012
  [arXiv:0910.5241 [hep-th]].


 \bibitem{BTZ} M. Ba\~{n}ados, C. Teitelboim, and J. Zanelli, Phys. Rev. Lett. 69 (1992) 1849 [arXiv:hep-th/9204099].
\bibitem{BH}
J. D.~Brown and  M.~Henneaux, Commun. Math. Phys.104:207-226,1986.
\bibitem{Henneaux:2009pw}
  M.~Henneaux, C.~Martinez and R.~Troncoso,
  Phys.\ Rev.\  D {\bf 79} (2009) 081502
  [arXiv:0901.2874 [hep-th]].
\bibitem{Anninos:2010pm}
  D.~Anninos, G.~Comp\`{e}re, S.~de Buyl, S.~Detournay and M.~Guica,
  arXiv:1005.4072 [hep-th].
  \bibitem{Gibbons} G. Gibbons and S.A. Hartnoll, Phys. Rev. D66 (2002) 064024
[arXiv:hep-th/0206202].
  \bibitem{Vish} C.V. Vishveshwara, Phys. Rev. D1 (1970) 2870.
\bibitem{Gregory} R. Gregory and R. Laflamme, Nucl. Phys. B428 (1994) 399 [arXiv:hep-th/9404071].
\bibitem{Teitelboim} M. Ba\~{n}ados, M. Henneaux, C. Teitelboim, and J. Zanelli, Phys. Rev. D48 (1993) 1506 [arXiv:gr-qc/9302012].

\bibitem{ivo}
  I.~Sachs,
  JHEP {\bf 0809} (2008) 073
  [arXiv:0807.1844 [hep-th]].
  \bibitem{Birtopbh} D. Birmingham, Class. Quant. Grav. 16 (1999) 1197
[arXiv:hep-th/9808032[.
\bibitem{BM} D. Birmingham and S. Mokhtari, Phys. Rev. D76 (2007) 124039 [arXiv:0709.2388 [hep-th]].



\bibitem{Kodama2} A. Ishibashi and H. Kodama, Prog. Theor. Phys. 110 (2003)901
[arXiv:hep-th/0305185].



\bibitem{KayWald} B.S. Kay and R.M. Wald, Class. Quant. Grav. 4 (1987) 893.

\bibitem{Abram} M. Abramowitz and I.A. Stegun, {\em Handbook of
Mathematical Functions}, Dover, New York, 1970.

\end{thebibliography}
\end{document}